\documentclass[conference, compsoc]{IEEEtran}
\ifCLASSINFOpdf
\else
\fi
\usepackage{times}
\usepackage{graphicx,amssymb,amsmath}

\newtheorem{thm}{Theorem}
\newtheorem{defn}[thm]{Definition}
\newcommand*{\metacomp}{\overrightarrow{\cdot}}

\begin{document}

\title{Formalizing Safety Requirements Using Controlling Automata}

\author{\IEEEauthorblockN{Zhe Chen\ $^\dagger$}
\IEEEauthorblockA{$^\dagger$\ Laboratory LATTIS, INSA, University of Toulouse\\
135 Avenue de Rangueil, 31077 Toulouse, France\\
Email: zchen@insa-toulouse.fr} \and \IEEEauthorblockN{Gilles Motet\
$^{\dagger,\ddagger}$}
\IEEEauthorblockA{$^\ddagger$\ Foundation for an Industrial Safety Culture\\
6 All\'ee Emile Monso, 31029 Toulouse, France\\
Email: gilles.motet@insa-toulouse.fr} }

\maketitle

\begin{abstract}
Safety is an important element of dependability. It is defined as
the absence of accidents. Most accidents involving
software-intensive systems have been system accidents, which are
caused by unsafe inter-system or inter-component interactions. To
validate the absence of system hazards concerning dysfunctional
interactions, industrials call for approaches of modeling system
safety requirements and interaction constraints among components.
This paper proposes such a formalism, namely interface control
systems (or shortly C-Systems). An interface C-System is composed of
an interface automaton and a controlling automaton, which formalizes
safe interactions and restricts system behavior at the meta level.
This framework differs from the framework of traditional model
checking. It explicitly separates the tasks of product engineers and
safety engineers, and provides a top-down technique for modeling a
system with safety constraints, and for automatically composing a
safe system that conforms to safety requirements. The contributions
of this work include formalizing safety requirements and a way of
automatically ensuring system safety.
\end{abstract}
\IEEEpeerreviewmaketitle

\section{System Safety Requirements}
Critical systems are always controlled by software applications,
which overcome the shortcomings of human control, but also introduce
new failure modes that are changing the nature of accidents
\cite{Lev04}. {\em Inter-system} and {\em inter-component}
dependability are becoming important, since industrials are
developing complicated software-intensive systems which consist of
numerous components (subsystems) and a huge number of actions (both
internal and interactive). A recent challenge of dependability is
the {\em system accident}, caused by increasing {\em inter-system}
and {\em inter-component couplings} and their {\em interactive
complexity} \cite{Lev08}\cite{Per99}. In contrast, accidents arising
from component failures are termed {\em component failure
accidents}.

{\em System safety} and {\em component reliability} are different
elements of dependability. They are system property and component
property, respectively \cite{Lev08}. {\em Reliability} is defined as
the capability that a component satisfies its specified behavioral
requirements, whereas {\em safety} is defined as the absence of
accidents --- events involving an unacceptable loss \cite{Lev95}.
People are now constructing intellectually unmanageable software
systems that go beyond human cognitive limits. This allows
potentially {\em unsafe interactions} to be undetected. Accidents
often result from hazardous interactions among perfectly functioning
components.

\begin{figure}
\centering
  \includegraphics[scale=0.4]{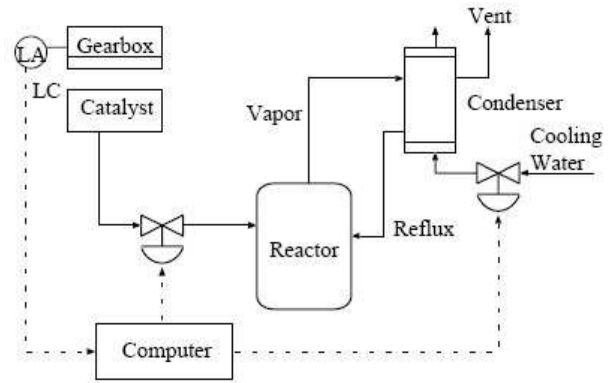}\\
  \caption{A Chemical Reactor Design}\label{Fig:Reactor}
\end{figure}

As an example, a system accident occurred in a batch chemical
reactor in England \cite{Kle82}. The design of the system is shown
in Fig. \ref{Fig:Reactor}. The computer controlled the input flow of
cooling water into the condenser and the input flow of catalyst into
the reactor by manipulating the valves. The computer was told that
if any component in the plant gets abnormal, it had to leave all
controlled variables as they were and to sound an alarm. On one
occasion, the computer just started to increase the cooling water
flow, after a catalyst had been added into the reactor. Then the
computer received an abnormal signal indicating a low oil level in a
gearbox, and it reacted as its requirements specified: sounded an
alarm and maintained all the control variables with their present
condition. Since the water flow was kept at a low rate, then the
reactor overheated, the relief valve lifted and the contents of the
reactor were discharged into the atmosphere.

Some other system accidents in avionics are also due to uncontrolled
interactions between components \cite{Lev01}. The self-destructing
explosion of Ariane 5 launcher was resulted from the successive
failures of the active Inertial Reference System (IRS) and the
backup IRS \cite{Lev01}. Ariane 5 adopted the same reference system
as Ariane 4. However, the profile of Ariane 5 was different from
that of Ariane 4 --- the acceleration communicated as input value to
IRS of Ariane 5 was higher. Furthermore, the {\em interactions
between IRS and other components} were not redefined and checked.
Due to the overflow of input value computation, the IRS stopped
working \cite{KT06}. Then, the signaled error was interpreted as a
launcher attitude, and led the control system to rotate the tailpipe
at the end stop \cite{GM02}.

In these accidents, the components are reliable in terms of
satisfying their specified requirements, but the systems are not
safe as a whole. Since most software related accidents have been
system accidents \cite{Lev04}, people need to model and constrain
interactions of system components to validate the absence of {\em
dysfunctional interactions}. As Leveson mentioned in STAMP
(Systems-Theoretic Accident Model and Processes) \cite{Lev04}, these
accidents result from inadequate {\em control} or enforcement of
{\em safety-related constraints} of the systems.

Traditionally, in order to validate the absence of system hazards,
industrials identify system safety requirements
\cite{AK01}\cite{Fir04}, and use model checking to verify if system
behaviors conform to safety requirements \cite{Bit01}.

In this paper, we will consider safety requirements as control
structures that restrict system behaviors at meta-model level. That
is, the two models of a system and its safety constraints will be
developed at the same time. Then the two models are ``combined'' to
deduce a safe system.

This paper is organized as follows: the architecture of our approach
is presented in Section 2. To illustrate the idea, a preliminary
introduction on interface automata appears in Section 3. The
interface C-System based on controlling automata is introduced in
Sections 4 and 5, where examples are used to illustrate how to
formalize safety rules and combine them with a system specification.
In Section 6, we compare our work to classic verification
techniques, such as model checking, and conclude the paper.

\section{The Architecture of our Methodology}
The most popular technique of system safety verification is {\em
model checking} \cite{CGP00}. Hundreds of checking patterns are
collected for system engineers \cite{DAC99} and specific uses in
safety engineering \cite{Bit01}. In this framework, we have three
steps in verifying a system. At first, we formalize system behavior
as a model (e.g., a finite-state transition system, a Kripke model
\cite{HR04}). At the second step, we specify the safety constraints
that we aim at validating using temporal logics \cite{DAC99}. At the
third step, a certain checking algorithm is used to search for a
counterexample which is an execution trace violating the specified
features. If the algorithm finds such a counterexample, we have to
modify the original design to ensure safety constraints.

Unlike model checking, our architecture takes another way. It
consists of the following steps:

(1) Modeling system behavior, including specifications of its
components, internal and external interactions, e.g., using
interface automata.

(2) Modeling system safety constraints using a certain formal
technique, e.g., controlling automata in this paper.

(3) Combining these two models to deduce a safe system model, that
is, a system model whose behavior is in accordance with its safety
constraints.

As \cite{Mot09} mentioned, system behavior specifies an {\em
operational semantics}, which defines what a system is able to do.
System behavior modeling is achieved by {\em product engineers}
(designers), such as programmers and developers. In the example of
the chemical reactor control system, the actions ``opening the
catalyst flow'', ``opening the cooling water flow'' and ``sounding
an alarm'' are actions of the system behavior.

In the second step, the model of safety constraints specifies a {\em
correctness semantics}, which defines what a system is authorized to
do. This process is the duty of {\em safety engineers} whose
responsibility is to assure system safety. Safety engineers may
consist of requirement engineers, testing engineers, managers from
higher socio-technical levels who define safety standards or
regulations \cite{Lev04}, etc. In the example of the chemical
reactor system, the constraint ``opening the catalyst flow must be
followed by opening the cooling water flow'' is an instance of
system safety constraints.

In the third step, in order to ensure system safety, we combine a
system model with its safety constraints model. Then we ensure that
the system is safe under the constraints specifying safety
requirements. A precondition of this approach is that we must
formalize safety requirements. And we also need to carefully define
the composition of a system model and its constraints model. We will
introduce such means based on controlling automata.

\section{Preliminary: Interface Automata}
To model component-based concurrent systems with different input,
output and internal actions, the theory of interface automata
\cite{AH01} extends Input/Output automata \cite{LT89}\cite{Lyn96},
which extends classic automata theory \cite{HU79}.

Unlike I/O automata, an interface automaton is not required to be
input-enabled (i.e., some inputs may be recognized as illegal in
some states) and only allows the composition of two automata (I/O
automata allow the composition of infinite automata), and a
synchronization of one output and one input action results a hidden
action after the composition.

\begin{defn}
An \textbf{interface automaton} (simply an automaton) is a tuple
$A=(Q,\Sigma^I,\Sigma^O,\Sigma^H,\delta,S)$, where:\\
(1) $Q$ is a set of \textbf{states}.\\
(2) $\Sigma^I,\Sigma^O,\Sigma^H$ are pairwise disjoint sets of
\textbf{input, output and internal actions}, respectively. Let
$\Sigma = \Sigma^I \bigcup \Sigma^O \bigcup \Sigma^H$ be the set of
\textbf{actions}.\\
(3) $\delta \subseteq Q \times \Sigma \times Q$ is a set of
\textbf{labeled transitions}.\\
(4) $S \subseteq Q$ is a set of \textbf{start states}, where
$|S|\leq1$. \hfill $\Box$
\end{defn}

In the graph notation, a transition $p_k:(q,a,q') \in \delta$ is
denoted by an arc from $q$ to $q'$ labeled $p_k:a$, where $p_k$ is
{\em the name of the transition}. To discriminate explicitly the
different sets of actions in diagrams, we may suffix a symbol ``?'',
``!'' or ``;'' to an input, output or internal action, respectively.

The composition of two composable automata allows the automata to
synchronize on shared actions, and asynchronously interleave all
other actions.

\begin{defn}
Two interface automata $A$ and $B$ are \textbf{composable} if
$\Sigma^H_A \cap \Sigma_B = \emptyset$, $\Sigma^I_A \cap \Sigma^I_B
= \emptyset$, $\Sigma^O_A \cap \Sigma^O_B = \emptyset$, $\Sigma^H_B
\cap \Sigma_A = \emptyset$. We let $shared(A,B)=\Sigma_A \cap
\Sigma_B$. \hfill $\Box$
\end{defn}

\begin{defn}
If $A$ and $B$ are composable interface automata, their
\textbf{product} $A \otimes B$ is the interface automaton defined
by\\
(1) $Q_{A \otimes B} = Q_A \times Q_B$\\
(2) $\Sigma^I_{A \otimes B} = (\Sigma^I_A \cup \Sigma^I_B) -
  shared(A,B)$\\
(3) $\Sigma^O_{A \otimes B} = (\Sigma^O_A \cup \Sigma^O_B) -
  shared(A,B)$\\
(4) $\Sigma^H_{A \otimes B} = \Sigma^H_A \cup \Sigma^H_B \cup
  shared(A,B)$\\
(5) \begin{tabbing}
             $\delta_{A \otimes B}$ \= $= \{$ \= $p_i:((v,u),a,(v',u))~|~p_i:(v,a,v')\in \delta_A$ \\
                                    \>        \> $\wedge a \not\in shared(A,B) \wedge u \in Q_B\}$\\
                                    \> $\cup~ \{p_j:((v,u),a,(v,u'))~|~p_j:(u,a,u')\in \delta_B$\\
                                    \>        \> $\wedge a \not\in shared(A,B) \wedge v \in Q_A\}$ \\
                                    \> $\cup~ \{p_{ij}:((v,u),a,(v',u'))~|~p_i:(v,a,v')\in \delta_A$\\
                                    \>        \> $\wedge p_j:(u,a,u')\in \delta_B \wedge a \in shared(A,B) \}$
        \end{tabbing}
(6) $S_{A \otimes B} = S_A \times S_B$. \hfill $\Box$
\end{defn}

Note that the name of the transition $p_{ij}$ of $A \otimes B$ may
contain the names of two original transitions $p_i \in \delta_A$ and
$p_j \in \delta_B$.

In the product $A \otimes B$, there may be \textbf{illegal states},
where one component is able to send an output $a \in shared(A,B)$
and the other is not able to receive $a$.

The composition of two interface automata $A,B$ is obtained by
restricting the product of the two automata to the set $Cmp(A,B)$ of
\textbf{compatible states}, which are the states from which there
exists a legal environment that can prevent entering illegal states.

\begin{defn}
If $A$ and $B$ are composable interface automata, their
\textbf{composition} $A || B$ is the interface automaton defined
by\\
(1) $Q_{A || B} = Cmp(A,B)$\\
(2) $\Sigma^I_{A || B} = \Sigma^I_{A \otimes B}$\\
(3) $\Sigma^O_{A || B} = \Sigma^O_{A \otimes B}$\\
(4) $\Sigma^H_{A || B} = \Sigma^H_{A \otimes B}$\\
(5) $\delta_{A || B} = \delta_{A \otimes B} \cap (Cmp(A,B) \times
\Sigma_{A || B} \times Cmp(A,B))$\\
(6) $S_{A || B} = S_{A \otimes B} \cap Cmp(A,B)$. \hfill $\Box$
\end{defn}

\section{Safety Constraints on a Single Component}
\begin{figure*}
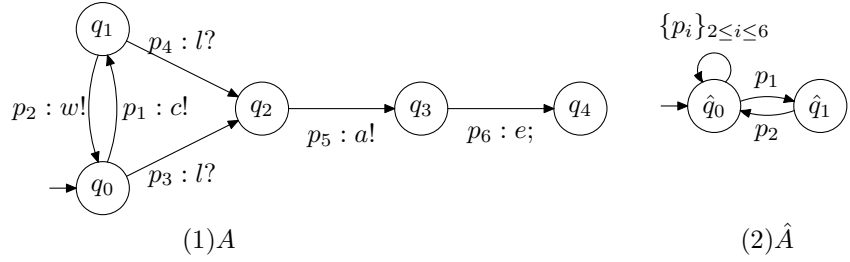

  \centering
  \includegraphics{fig.1}\ \ \ \ \ \includegraphics{fig.2}\\
  \caption{Automata of the Reactor Control System}\label{Fig:DFA_RCS}
\end{figure*}

In this section, we start from a simple case -- modeling safety
constraints on a single component. In the example of the batch
chemical reactor (C.f. Fig. \ref{Fig:Reactor}), the computer system
behavior is modeled using an interface automaton $A$ of Fig.
\ref{Fig:DFA_RCS}(1). The automaton $A$ includes a set of input
actions $\Sigma^I = \{l\}$ (low oil signal), a set of output actions
$\Sigma^O = \{c,w,a\}$ (opening catalyst flow, opening water flow,
sounding an alarm, respectively), and a set of internal actions
$\Sigma^H=\{ e \}$ (ending all operations).

The normal operational behavior includes opening the catalyst flow
($p_1$), then opening the water flow ($p_2$), etc., resulting in an
infinite execution trace $p_1p_2p_1p_2...$. To respond to abnormal
signals as soon as possible, the states $q_0,q_1$ both have a
transition labeled $l$, which leads to a state that can sound an
alarm ($p_5$) and stop the process ($p_6$). Unfortunately, this
design leads to hazardous behaviors: $(cw)^* clae$, that is, after a
sequence of opening catalyst and water flows $(cw)^*$, then the
catalyst flow is opened ($c$) when an abnormal signal is received
($l$), then an alarm is sounded ($a$). So water is not added after
the catalyst flow is opened. This sequence of events leads to the
accident mentioned in Section 1.

Note that this hazard is due to the uncontrolled sequences of
transitions --- $p_1$ must be followed by $p_2$ and not by $p_4$. To
solve this problem, we need to specify the authorized sequences
(satisfying safety constraints) on the transitions $\delta$ and not
on the actions $\Sigma$. Thus, these constraints are not at the
behavioral model level, but at the meta-model level. We propose the
concept of {\em controlling automata} to formalize safety
constraints. Then, we combine a controlling automaton with the
system automaton.

\begin{defn}
A \textbf{controlling automaton} $\hat{A}$ over an interface
automaton $A=(Q,\Sigma,\delta,S)$ is a tuple
$\hat{A}=(\hat{Q},\hat{\Sigma},\hat{\delta},\hat{S})$, where:\\
(1) $\hat{Q}$ is a set of \textbf{states} disjoint with $Q$.\\
(2) $\hat{\Sigma}$ is a set of \textbf{terminals}, such that
    $\hat{\Sigma}=\delta$.\\
(3) $\hat{\delta} \subseteq \hat{Q} \times \hat{\Sigma}
\times \hat{Q}$ is a set of \textbf{labeled transitions}.\\
(4) $\hat{S} \subseteq \hat{Q}$ is a nonempty set of \textbf{start
states}. \hfill $\Box$
\end{defn}

\begin{figure}[htb]
  \centering
  \includegraphics{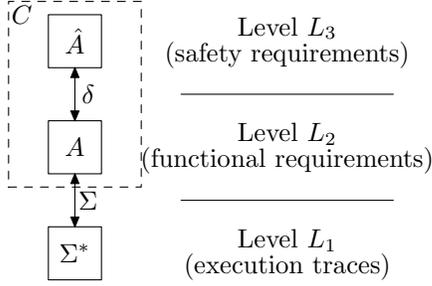}\\
  \caption{A 3-levels Overview}\label{Fig:3_levels}
\end{figure}

Note that the transitions $\delta$ of $A$ are terminals of
$\hat{A}$, so we say that $\hat{A}$ is at the meta level of $A$.
Figure \ref{Fig:3_levels} illustrates the 3 levels in our framework.
Let $\Sigma^*$ be a set of execution traces of actions, $A$
describes the behavior on $\Sigma$. $\hat{A}$ specifies the behavior
on the $A$-transitions ($\hat{\Sigma}=\delta$), that is, a behavior
on the behavior of $A$. This meta-behavior expresses safety
requirements.

In the example, to prevent accidents, we need to impose the safety
constraint ``opening catalyst must be followed by opening water,''
that is, ``whenever the transition $p_1:c$ occurs, the transition
$p_2:w$ must occur after that''. This constraint can be formalized
as a controlling automaton $\hat{A}$ of Fig. \ref{Fig:DFA_RCS}(2).
When we express this constraint, we only specify the sequence of
transitions $p_1,p_2$ at the meta-model level, and we concern little
about the implementation of the system at the model level. The next
step is to compose the system automaton $A$ with its controlling
automaton $\hat{A}$, and automatically generate a system $C$
satisfying the safety requirement.

\begin{defn}
The \textbf{meta-composition} $C$ of an interface automaton
$A=(Q,\Sigma,\delta,S)$ and a controlling automaton
$\hat{A}=(\hat{Q},\hat{\Sigma},\hat{\delta},\hat{S})$ over $A$ is a
tuple:
\begin{equation}
C = A \metacomp \hat{A} = (Q \times \hat{Q}, \Sigma, \delta', S
\times \hat{S})
\end{equation}
where $p_k: ((q_i,\hat{q}_j), a, (q_m,\hat{q}_n)) \in \delta'$ iff,

(1) $p_k:(q_i,a,q_m) \in \delta$, and

(2) $(\hat{q}_j, p_k, \hat{q}_n) \in \hat{\delta}$.

We say that $A$ and $\hat{A}$ constitute an \textbf{interface
control system} (or simply \textbf{interface C-System}). \hfill
$\Box$
\end{defn}

The symbol $\metacomp$ is called {\em meta-composition operator},
and read ``meta-compose''. Its left and right operands are an
automaton and a controlling automaton, respectively. Notice that an
interface C-System is equivalent to the meta-composition $C$ of an
interface automaton and a controlling automaton.

\begin{figure}[htb]
  \centering
  \includegraphics{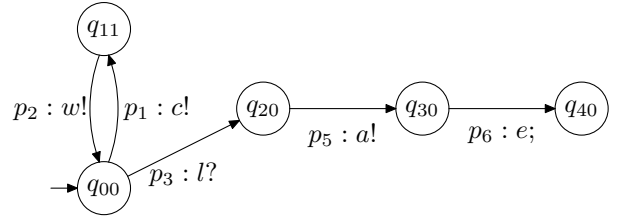}\\
  \caption{The Meta-Composition $C$ } \label{Fig:DFA_RCS_C}
\end{figure}

Notice that $\delta = \{p_k\}_{k \in \mathcal{K}}$ plays a key role
in associating transitions of $A$ and terminals of $\hat{A}$. For
our example, we combine the automata $A$ and $\hat{A}$ of Fig.
\ref{Fig:DFA_RCS}, thus we get the automaton $C = A \metacomp
\hat{A}$ of Fig. \ref{Fig:DFA_RCS_C} where $q_{ij}$ denotes
$(q_i,\hat{q}_j)$.

The meta-composition contains exactly all the paths satisfying the
safety constraint. Formally, we have the following theorem (the
proof is omitted for its simpleness and intuitiveness from the
definition):

\begin{thm}
Given $A,\hat{A}$ and the meta-composition $C$, an \textbf{execution
trace} $t_\Sigma \in \Sigma^*$ is recognized by $C$ iff, $t_\Sigma$
is recognized by $A$, and its \textbf{transition trace} $t_\delta
\in \delta^*$ is recognized by $\hat{A}$. \hfill $\Box$
\end{thm}

Obviously, the set of traces of $C$ is a subset of the traces of
$A$. Formally, let $L(A)$ be the set of traces of $A$ (i.e. the
language of $A$), we have $L(C) \subseteq L(A)$.

Thanks to $\hat{A}$, the hazardous execution traces, for example
$cwclae$, which exists in $A$, will be eliminated, because its
transition trace $p_1p_2p_1p_4p_5p_6 \not\in L(\hat{A})$ (the
language of $\hat{A}$). The comparison between $A$ of Fig.
\ref{Fig:DFA_RCS}(1) and $C$ of Fig. \ref{Fig:DFA_RCS_C} highlights
the hazardous transition $p_4$ of $A$. However, in general, this
diagnosis is much more complex and cannot be achieved manually,
since a real system $A$ has too many states to be expressed clearly
on a paper. That is why we developed a formal and automated method
for eliminating hazardous transitions.

\section{Safety Constraints on Multi-Components}

\begin{figure*}
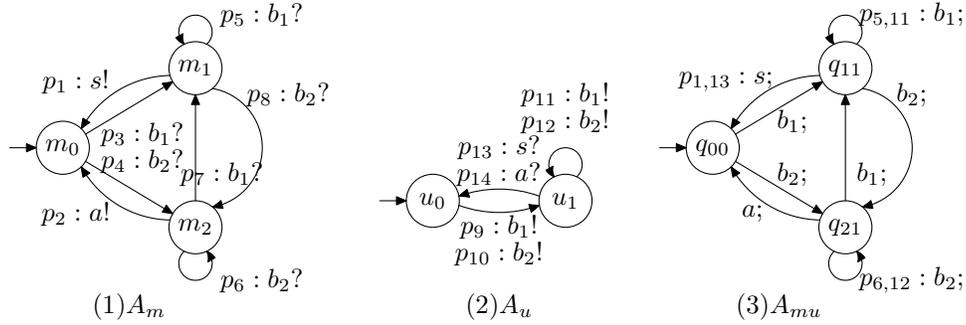

  \centering
  \includegraphics{fig.4}\ \ \ \ \ \includegraphics{fig.5}\ \ \ \ \ \includegraphics{fig.6}\\
  \caption{Automata of the Candy Machine System}\label{Fig:DFA_CM}
\end{figure*}

To illustrate this case, we use an example concerning a system
composed of two components with interactions: a candy vending
machine and a customer. We hope that, since this class of examples
is so popular in the literatures of formal methods (e.g., Hoare's
Communicating Sequential Processes (CSP) and I/O automata
\cite{LT89}), they will provide an interesting illustration of our
idea. The candy machine $A_m$, specified in Fig.
\ref{Fig:DFA_CM}(1), may receive inputs $b_1,b_2$ indicating that
buttons 1 and 2 are pushed, respectively. It may output $s,a$,
indicating candy dispensation actions, SKYBARs and ALMONDJOYs,
respectively. The machine may receive several inputs before
delivering a candy. A greedy user $A_u$, specified in Fig.
\ref{Fig:DFA_CM}(2), can push buttons $b_1,b_2$ or get a candy
$s,a$. The greedy user does not wait for a candy bar before pressing
a button again.

The composition of the machine behavior and the user behavior is
defined by $A_{mu}=A_m || A_u$ of Fig. \ref{Fig:DFA_CM}(3), where
$q_{ij}$ denotes the composite state $(m_i,u_j)$, $p_{i,j}$ denotes
two synchronized transitions $\{ p_{i}, p_{j} \}$. A transition of
the composition may be composed of two transitions of components.
For example, $p_{1,13}:s$ is a synchronization of $p_1:s!$ and
$p_{13}:s?$, which belong to $A_m$ and $A_u$, respectively.
Generally, a transition of $A =P || Q$ may be composed of one or two
transitions of its components, where two transitions constitute a
synchronization.

In the context of meta-composition, a composite transition is
allowed if and only if both of its sub-transitions are allowed by
its controlling automaton. Thus, we define the meta-composition
operator as follows:

\begin{defn}
The \textbf{meta-composition} (or interface C-System) $C$ of a
composition $A = P || Q$ and a controlling automaton
$\hat{A}=(\hat{Q},\hat{\Sigma},\hat{\delta},\hat{S})$ over $A$ is a
tuple:
\begin{equation}
C = A \metacomp \hat{A} = ( Q_A \times \hat{Q}, \Sigma_A, \delta',
S_A \times \hat{S})
\end{equation}
where $p_{\mathcal{I}}: ((v,u,q), a, (v',u',q')) \in \delta'$
($p_{\mathcal{I}}$ contains a set of transitions $\{p_k\}_{k \in
\mathcal{I}}$) iff,

(1) $p_{\mathcal{I}}: (((v,u), a, (v',u')) \in \delta_A$, and

(2) $\forall k: k \in \mathcal{I} \bullet (q,p_k,q') \in
\hat{\delta}$. \hfill $\Box$
\end{defn}

Notice that the specification of the example allows a hazardous
situation: the greedy user repeatedly pushes the buttons without
giving the machine a chance to dispense a candy bar (e.g., the
transition labeled $p_{5,11}:b_1$ of $q_{11}$ does not allow the
transition $(q_{11}, s, q_{00})$ to be fired). To prevent this
situation, the following constraints forbid successive occurrences
of pressing buttons: ``the transitions $p_{11},p_{12}$ are not
allowed, when interactions occur between the machine and the user''.
Differing from the previous example, this type of constraints needs
to synchronize the actions of the machine and of the user.

\begin{figure*}
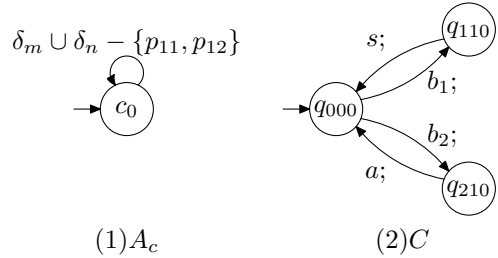

  \centering
  \includegraphics{fig.7}\ \ \ \ \includegraphics{fig.8}\\
  \caption{A Safety Constraint of the Candy Machine System}\label{Fig:DFA_CM_C}
\end{figure*}

Formalizing the constraints, the semantics of the controlling
automaton $A_c$ of Fig. \ref{Fig:DFA_CM_C}(1) is: whenever the user
pushes a button ($p_9,p_{10}$), she or he cannot push it again
($p_{11},p_{12}$), but can only wait for a candy bar.

Combining the whole system $A_{mu}$ with its constraint $A_c$, we
get the system $C = (A_m || A_u) \metacomp A_c$ in Fig.
\ref{Fig:DFA_CM_C}(2), where $q_{ijk}$ denotes the composite state
$(m_i,u_j,c_k)$. All of its execution traces satisfy the constraint,
and thus prevent the hazardous situation.

Since we formally defined the {\em meta-composition operator}, it
can be easily implemented to be an automated tool. Thus, it can be
applied to more complex systems.

\section{Conclusion}
We proposed formalizing system safety requirements using controlling
automata. As we illustrated using examples, this approach can
formally model safe interactions between components or systems. This
framework differs from the one of model checking. It explicitly
separates the tasks of product engineers and safety engineers, and
provides a technique for modeling a system with safety constraints,
and for automatically composing a safe system that conforms to
safety requirements.

The essential ideas of our approach are the separation and
formalization of the system specification $A$ (core functional
requirements) and the safety constraints $\hat{A}$ (safety
requirements). The automaton $A$ handles inputs to produce outputs
using activities depending on the states, whereas the controlling
automaton $\hat{A}$ treats activities to produce the set of
acceptable activities depending on safety requirements.

Our framework has different objectives and uses different approaches
to those of model checking. Model checking techniques use {\em a
bottom-up approach} --- it verifies execution traces $\Sigma^*$ at
the lower level $L_1$ to prove the correctness and safety of the
system model $A$ at the middle level $L_2$ (see Fig.
\ref{Fig:3_levels}). However, our proposal uses {\em a top-down
approach} --- we model safety requirements as acceptable sequences
of transitions ($\delta^*$) at the higher level $L_3$ to ensure the
correct use of $A$. Then any execution trace (at $L_1$) that
conforms to the meta-composition $C$ is definitely a safe execution.
The two techniques are complementary. Model checking may be used to
reduce the design fault likelihood, and our approach can be applied
to avoid behavior that are not in accordance with some critical
safety requirements.

This paper continues our work on {\em C-Systems} (formal language
control systems). In \cite{CM09a}, we actually proposed the {\em
input/output C-System}. The {\em context-free C-System} was proposed
in \cite{CM09c} for restricting the use of modeling languages, in
order to ensure guidelines and consistency rules of UML.

In the future, it might be a good direction to study the
formalization of parameterized safety constraints. Another direction
is empirical case study on applying this formalism in large and
complex systems.

\bibliographystyle{IEEEtran}
\bibliography{ref}

\end{document}